\documentstyle[12pt]{article}
\begin{document}


\sloppy
\title
{\hfill{\normalsize\sf FIAN/TD/01-01}    \\
            \vspace{1cm}
{\Large The general form of the star--product on the Grassman algebra}
}
\author
 {
       I.V.Tyutin
          \thanks
             {E-mail: tyutin@lpi.ru}
  \\
               {\small \phantom{uuu}}
  \\
           {\it {\small} I.E.Tamm Department of Theoretical Physics,}
  \\
               {\it {\small} P.N.Lebedev Physical Institute,}
  \\
         {\it {\small} 117924, Leninsky Prospect 53, Moscow, Russia.}
 }
\date{ }
\maketitle
\begin{abstract}
We study the general form of the noncommutative associative product
(the star--product) on the Grassman algebra; the star--product is treated as
a deformation of the usual "pointwise" product. We show that up to a
similarity transformation, there is only one such product. The relation of
the algebra ${\cal F}$, the algebra
of elements of the Grassman algebra with the star--product as a product,
to the Clifford algebra is discussed.
\end{abstract}

\newpage

\section{Introduction}

The noncommutative associative product of functions on a phase space (the
star--product, the $*$--product in what follows) arises frequently in the
physical literature. The traditional region where the $*$--product
construction is conventionally used is the problem on quantizing the
classical theory (i.e., the Poisson bracket) in the case where the phase
space is a nontrivial manifold, the so--called geometrical or deformation
quantizations (see \cite {KarMas} and references wherein). The
relations of the $*$--product to quantum groups were recently found
\cite {FlaSte}, \cite {BonFlaPin}. The gauge theories on the noncommutative
spaces, the so--called noncommutative gauge theories,  are formulated in
terms of the $*$--product (see \cite {Nek} and \cite {AS} and references
wherein).

The structure and the properties of the $*$--product on the usual (even)
manifolds are
investigated in details \cite{KarMas}, \cite{Fed3} and \cite{Kon}. The
BRST formulation of the Fedosov construction \cite{Fed2} for the
$*$--product on symplectic manifolds was recently found \cite{GriLya}. On
the other hand, the $*$--product on supermanifolds is not sufficiently
studied.  It is easy to  extend formally the noncommutative product proposed
in \cite{Gro} to the supercase \cite{Ber} (see also \cite{BatFra},
\cite{FraLin} \cite{GozReu1}), however, a discussion of the uniqueness of
the $*$--product (the uniqueness of the "pointwise" product deformation)
is absent in the
literature. In the present paper, we investigate the general form of the
associative $*$--product, treated as a deformation of the "pointwise"
product, on the Grassman algebra of a finite number of generators. We
show that up to a similarity transformation, only one such product
exists.

The paper is organized as follows. In Sec. 2, we describe the formulation of
the
problem. In Sec. 3, we present the solution of the associativity equation for
the $*$--product. In Sec. 4, we prove the
uniqueness (up to a similarity transformation) of the $*$--product
and formulate the basic results. In Sec. 5, the
relation of the algebra ${\cal F}$ of elements of the Grassman algebra with
$*$--product as a product to the Clifford algebra is discussed.

{\bf Notation and conventions.}

$\xi^\alpha$, $\alpha=1,\ldots,n$ are odd anticommuting generators
of the Grassman algebra:
$$
\varepsilon(\xi^\alpha)=1,\quad
\xi^\alpha\xi^\beta+\xi^\beta\xi^\alpha=0,\quad\partial_\alpha\equiv
{\partial\over\partial\xi^\alpha},
$$
$\varepsilon(A)$ denotes the Grassman parity of $A$;
$$
[\xi^\alpha]^0\equiv1,\quad[\xi^\alpha]^k\equiv\xi^{\alpha_1}
\cdots\xi^{\alpha_k}, 1\le k\le n,\quad
[\xi^\alpha]^k=0, k>n,
$$
$$
[\partial_\alpha]^0\equiv1,\quad[\partial_\alpha]^k\equiv
\partial_{\alpha_1}\cdots\partial_{\alpha_k},1\le k\le n, \quad
[\partial_\alpha]^k=0, k>n,
$$
$$
[\overleftarrow\partial_\alpha]^k\equiv\overleftarrow\partial_{\alpha_1}
\cdots\overleftarrow\partial_{\alpha_k},
$$

$$
T_{\ldots[\alpha]_k\ldots}\equiv T_{\ldots\alpha_1\ldots\alpha_k\ldots},
\quad T_{\ldots\alpha_i\alpha_{i+1}\ldots}=-
T_{\ldots\alpha_{i+1}\alpha_i\ldots},\quad i=1,\ldots,k-1.
$$

\section{Setting the problem}

We consider the Grassman algebra ${\cal G}_K$ over the field $K$ $=$ ${\bf
C}$ or ${\bf R}$ with the generators $\xi^\alpha$, $\alpha=1,2,\ldots,n$. The
general element $f$ of the algebra (a function of the generators) is
$$
f\equiv f(\xi)=\sum\limits_{k=0}^n{1\over k!}f_{[\alpha]_k}[\xi^\alpha]^k,
\quad f_{[\alpha] _0}\equiv f_0,f_{[\alpha] _k}\in K,\quad\forall k.
$$

We also introduce the basis $\{e^I\}$ in the algebra ${\cal G}_K$:
\begin{equation}\label{Bas}
\begin{array}{c}
I=0,\,\alpha_i,\,\alpha_{i_1}\alpha_{i_2},\,\ldots,\,
\alpha_{i_1}\!\cdots\alpha_{i_k},\,\ldots\,,\quad i_1<i_2<\cdots<i_k, \\
e^0=1,\quad e^{\alpha_{i_1}\alpha_{i_2}\cdots\alpha_{i_k}}=
\xi^{\alpha_{i_1}}\xi^{\alpha_{i_2}}\cdots\xi^{\alpha_{i_k}},
\quad k=0,1, \ldots, n.
\end{array}
\end{equation}
The basis contains $2^n$ elements. Any element $f$ of the algebra
${\cal G}_K$ can be uniquely represented as
\begin{equation}\label{Rep}
f=\sum\limits_If_Ie^I,\quad f_I\in K.
\end{equation}

The integral over the generators is normalized as follows:
$$
\int d\xi[\xi^\alpha]^k=0,\,k\le n-1,\quad\int d\xi[\xi^\alpha]^n=
\varepsilon^{[\alpha]_n},\quad\varepsilon(d\xi)=n\,(mod\,\,2).
$$
$$
\int d\xi^\prime\delta(\xi-\xi^\prime)f(\xi^\prime)=f(\xi),\quad\varepsilon
(\delta(\xi))=n\,(mod\,\,2).
$$

We are interested in the general form of the associative even $*$--product
for arbitrary elements $f_1,f_2$:
$$
f_3(\xi) =f_1*f_2 (\xi)\equiv P(\xi|f_1, f_2),
$$
i.e., the general form of the bilinear mapping ${\cal G} _K\times{\cal
G}_K\rightarrow{\cal G}_K $ satisfying the associativity equation:
\begin{equation}\label{AE}
f_1 * (f_2*f_3)=(f_1*f_2)*f_3,
\end{equation}
and having the parity property
$$
\varepsilon(f_1*f_2)=\varepsilon(f_1)+\varepsilon(f_2)\,(mod\,\,2).
$$

We solve associativity equation (\ref{AE}) in terms of the series
in the deformation parameter $\hbar$:
\begin{equation}\label{m1Ser}
*=\sum\limits_{k=0}\hbar^k*_k,
\end{equation}
where the $*_0$--product, that is the boundary condition, is the usual
"pointwise" product:
\begin{equation}\label{m1BC}
f_1*_0f_2(\xi)=f_1(\xi)f_2(\xi).
\end{equation}
Thus, we search for the general form of a possible even associative
deformation of the "pointwise" product on the Grassman algebra.

It is easy to verify that the $*$--product (as any bilinear mapping) can
be written in the coordinate representation form:
\begin{equation}\label{CR}
f_3(\xi)=\int\!\!d\xi_2d\xi_1P(\xi|\xi_1,\xi_2)f_1(\xi_1)f_2(\xi_2),\quad
\varepsilon(P(\xi|\xi_1,\xi_2))=0
\end{equation}
with a certain function (the kernel) $P(\xi|\xi_1,\xi_2)$, or in the
momentum representation form:
$$
f_3(\xi)=f_1(\xi)\!\!\!\sum_{k_1,k_2=0}^n[\overleftarrow\partial_\beta]^{k_1}
P^{[\beta]_{k_1}|[\gamma]_{k_2}}(\xi)[\partial_\gamma]^{k_2}f_2(\xi),\quad
\varepsilon(P^{[\beta]_{k_1}|[\gamma]_{k_2}}(\xi))=k_1+k_2\,(mod\,\,2)
$$
with certain coefficient functions $P^{[\beta]_{k_1}|[\gamma]_{k_2}} (\xi)$.

Let some associative even $*$--product exist, and let $T$ be the nonsingular
even linear mapping ${\cal G}_K\rightarrow{\cal G}_K$:
$$
Tf(\xi)\equiv T(\xi|f)=\int\!\!d\xi_1T(\xi|\xi_1)f(\xi_1)=\sum_{k=0}^n
T^{[\alpha]_k}(\xi)[\partial_\alpha]^kf(\xi),
$$
$$
\varepsilon(T(\xi|\xi_1))=n\,(mod\,\,2),\quad
\varepsilon(T^{[\alpha]_k}(\xi))=k\,(mod\,\,2).
$$
Then the $* _T$--product
$$
f_1 * _ Tf_2 (\xi) \equiv T ^ {-1} \biggl (Tf_1*Tf_2\biggr) (\xi)
$$
is also an associative even $*$--product. We say that the
$*_T$--product and the $*$--product are related by a ($T$--)similarity
transformation, or ($T$--)equivalent. We note that any nonsingular even
change of the algebra generators:
\begin{equation}\label{m2Change}
T_\eta\xi^\alpha\!=\!\eta^\alpha(\xi),\,T_\eta^{-1}\xi^\alpha\!=\!
T_\zeta\xi^\alpha\!=\!\zeta^\alpha(\xi),\quad T_\eta f(\xi)=f(\eta(\xi)),
\,T_\eta^{-1}f(\xi)=T_\zeta f(\xi)=f(\zeta(\xi)),
\end{equation}
where $\zeta^\alpha(\xi)$ is the inverse change,
$\eta^\alpha(\zeta(\xi))=\xi^\alpha$,
$\varepsilon(\eta^\alpha)=\varepsilon(\zeta^\alpha)=1$,
induces the similarity transformation. Any nonsingular even change can be
represented as
$$
T=e^{u^\alpha(\xi)\partial_\alpha}
$$
with some functions $u^\alpha(\xi)$, $\varepsilon(u^\alpha(\xi))=1$.

The associativity equation for the first order deformation $*_1$ is:
\begin{equation}\label{m13}
\biggl(f_1*_1(f_2f_3)\biggr)(\xi)+f_1(\xi)\biggl(f_2*_1f_3\biggr)(\xi)-
\biggl(f_1*_1f_2\biggr)(\xi)f_3(\xi)-\biggl((f_1f_2)*_1f_3\biggr)(\xi)=0.
\end{equation}
The trivial solution of eq. (\ref{m13}) is $*_{1triv}$ that is the
first order approximation to the $*_{0T}$--product $T$--equivalent to the usual
product with $T=I+\hbar t_1+O(\hbar^2)$, $I$ denotes
the identical mapping. This solution is of the form
\begin{equation}\label{m14}
f_1*_{1triv}f_2=(t_1f_1)f_2+f_1(t_1f_2)-t_1(f_1f_2).
\end{equation}
The problem is to describe all nontrivial solutions of eq. (\ref{m13}), which
appears to be sufficient for describing all solutions of associativity
equation (\ref{AE}).

Consider the algebra ${\cal A}=\sum\limits_\oplus{\cal A}_k$ of even
$k$--linear functionals $\Phi_k(\xi|f_1,\ldots,f_k)\in{\cal A}_k$
(of $k$--linear mappings $({\cal G} _K\times)^k$ $\rightarrow$ ${\cal G}_K $),
$\varepsilon(\Phi_k(\xi|f_1,\ldots,f_k))=\varepsilon(f_1)+\cdots+
\varepsilon(f_k)$ $(mod\,\,2)$, $k\ge1 $, ${\cal A}_0=K$, with
the (noncommutative) product (the mapping
${\cal A}_k\times{\cal A}_p\rightarrow{\cal A}_{k+p}$)
$$
\Phi_k\Psi_p(\xi|f_1,\ldots,f_{k+p})=\Phi_k(\xi|f_1,\ldots,f_k)
\Psi_p (\xi|f_{k+1},\ldots,f_{k+p}).
$$
In this algebra, there exists the natural grading $g$, $g(\Phi_k)=k$,
turning the algebra ${\cal A}$ into the graded algebra, $g({\cal A}_k)=k$,
and the linear operator $d_H$, the differential coboundary Hochschild operator
${\cal A}_k\rightarrow{\cal A}_{k+1}$, $g(d_H)=1$, acting according to the
rule:
$$
d_H\Phi_k(\xi|f_1,\ldots,f_{k+1})=f_1(\xi)\Phi_k(\xi|f_2,\ldots,f_{k+1})+
$$
$$
+\sum\limits_{i=1}^k(-1)^i
\Phi_k(\xi|f_1,\ldots,f_{i-1},f_if_{i+1},f_{i+2},\ldots,f_{k+1})+
(-1)^{k+1}\Phi_k(\xi|f_1,\ldots,f_k)f_{k+1}(\xi),
$$
$$
d_H {\cal A} _0=0.
$$
It is easy to verify that the operator $d_H$ is a graded odd
differentiation,
$$
d_H(\Phi_k\Psi_p)=(d_H\Phi_k)\Psi_p+(-1)^{g(\Phi_k)}\Phi_kd_H\Psi_p,
$$
and nilpotent,
$$
d_H^2=0.
$$

Equations (\ref{m13}) and (\ref{m14}) are written in terms of the operator
$d_H$ as
\begin{equation}\label{m15}
d_H*_1=0,
\end{equation}
\begin{equation}\label{m16}
*_{1triv}=d_Ht_1.
\end{equation}
This means that $*_1$ belongs to the second Hochschild cohomology
group, while the first order trivial deformations of the "pointwise"
product are coboundaries, i.e., they belong to the zero Hochschild cohomology.

\section{Solution of the associativity equation}

We find the general solution of eq. (\ref{m13}) for the first order
in $\hbar$ deformation of the "pointwise" product. We follow the method in
\cite{BT8}. It is convenient to use coordinate representation (\ref{CR}).
Associativity equation (\ref{AE}) for the $*$--product is reduced to the
equation for the kernel:
\begin {equation}\label{m2Ass}
\int d\xi_4P(\xi|\xi_4,\xi_3)P(\xi_4|\xi_1,\xi_2)=(-1)^n
\int d\xi_4P(\xi|\xi_1,\xi_4)P(\xi_4|\xi_2,\xi_3).
\end{equation}

With the boundary condition taken into account, the kernel of the $*$--product
has the following power series expansion in $\hbar$:
$$
P(\xi|\xi_1,\xi_2)=\delta(\xi-\xi_1)\delta(\xi-\xi_2)+
\hbar m_1(\xi|\xi_1,\xi_2)+O(\hbar^2).
$$
The equation for the kernel $m_1$ follows from eq. (\ref{m2Ass}):
\begin{equation}\label{m2Ass'}
\delta(\xi_1)\tilde{m}_1(\xi|\xi_2,\xi_3)+
\delta(\xi_3-\xi_2)\tilde{m}_1(\xi|\xi_1,\xi_3)=
\delta(\xi_3)\tilde{m}_1(\xi|\xi_1,\xi_2)+
\delta(\xi_1-\xi_2)\tilde{m}_1(\xi|\xi_1,\xi_3),
\end {equation}
$$
\tilde{m}_1(\xi|\xi_1,\xi_2)\equiv m_1(\xi|\xi_1+\xi,\xi_2+\xi).
$$
Multiplying eq. (\ref{m2Ass'}) by $\xi_1^\alpha$ and integrating over
$d\xi_1$, we obtain
\begin{equation}\label{m2Rel}
\xi_1^\alpha\tilde{m}_1(\xi|\xi_1,\xi_2)=
[(-1)^n\delta(\xi_1-\xi_2)-\delta(\xi_2)]a_1^\alpha(\xi|\xi_1),
\end{equation}
$$
a_1^\alpha(\xi|\xi_1)=\int d\xi_2\xi_2^\alpha\tilde{m}_1(\xi|\xi_2,\xi_1),
\quad\varepsilon(a^\alpha_1)=n+1\,(mod\,\,2).
$$
Multiplying relation (\ref{m2Rel}) by $\xi^\beta_1$ and taking
the antisymmetry in $\alpha$ and $\beta$ of the l.h.s. of the resulting
relation  into account, we find
$$
[(-1)^n\delta(\xi_1-\xi_2)-\delta(\xi_2)](\xi^\alpha_1a_1^\beta(\xi|\xi_1)+
\xi^\beta_1a_1 ^\alpha (\xi |\xi_1)) =0,
$$
whence it follows
\begin{equation}\label{m2Rel'}
\xi^\alpha_1a_1^\beta(\xi|\xi_1)+\xi^\beta_1a_1^\alpha(\xi|\xi_1)
=-2\delta(\xi_1)\omega_1^{\alpha\beta}(\xi),
\end{equation}
with a certain function $\omega_1^{\alpha\beta}(\xi)$,
$\omega_1^{\alpha\beta}=\omega_1^{\beta\alpha}$,
$\varepsilon(\omega^{\alpha\beta}_1)=0$. We represent the function
$a^\alpha_1(\xi |\xi_1)$ as
$$
a^\alpha_1(\xi|\xi_1)=-\omega_1^{\alpha\beta}(\xi)
{\partial\over\partial\xi^\beta_1}\delta(\xi_1)+
\bar{a}^\alpha_1(\xi|\xi_1),
$$
It follows from eq. (\ref{m2Rel'}) that
$$
\xi^\alpha_1\bar{a}_1^\beta(\xi|\xi_1) +
\xi_1^\beta\bar{a}_1^\alpha(\xi|\xi_1)=0,
$$
from where we find
$$
\bar{a}_1^\alpha(\xi|\xi_1)=\xi_1^\alpha\bar{a}_1(\xi|\xi_1),\quad
\varepsilon(\bar{a}_1)=n\,(mod\,\,2).
$$
Substituting the obtained representation for $a_1^\alpha$ into relation
(\ref{m2Rel}), we obtain:
$$
\xi_1^\alpha\tilde{m}_1(\xi|\xi_1,\xi_2)=\xi_1^\alpha[\delta(\xi_1-\xi_2)
\bar{a}_1(\xi|\xi_1)-(-1)^n\delta(\xi_2)\bar{a}_1(\xi|\xi_1)+
\delta(\xi_1){\overleftarrow\partial\over\partial\xi_1^\beta}
\omega^{\beta\gamma}_1(\xi){\partial\over\partial\xi^\gamma_2}\delta(\xi_2)],
$$
whence it follows
$$
 \tilde{m}_1(\xi|\xi_1,\xi_2)=\delta(\xi_1-\xi_2)
\bar{a}_1(\xi|\xi_1)-(-1)^n\delta(\xi_2)\bar{a}_1(\xi|\xi_1)+ \delta (\xi_1)
\bar {b} _1 (\xi |\xi_2) +
\delta(\xi_1){\overleftarrow\partial\over\partial\xi_1^\beta}
\omega^{\beta\gamma}_1(\xi){\partial\over\partial\xi^\gamma_2}\delta(\xi_2)],
$$
with a certain function $\bar{b}_1(\xi|\xi_2)$,
$\varepsilon(\bar {b}_1)=n\,(mod\,\,2)$.

Thus, the first order deformation of the "pointwise" product can be
represented in the form
$$
f_1*_1f_2(\xi)=m_1(\xi|f_1,f_2)=
$$
$$
f_1(\xi)\overleftarrow{\partial}_\alpha
\omega_1^{\alpha\beta}(\xi)\partial_\beta f_2(\xi)+f_1(\xi)t_1(\xi|f_2)
-t_1(\xi|f_1f_2)+t_1(\xi|f_1)f_2(\xi)+c_1(\xi|f_1f_2)-c_1(\xi|f_1)f_2(\xi),
$$
where we introduce the notation $t_1(\xi|\xi_1)=\bar{b}_1(\xi|\xi_1-\xi)$,
$c_1(\xi|\xi_1)=\bar{a}_1(\xi|\xi_1-\xi)+\bar{b}_1(\xi|\xi_1-\xi)$,
$t_1(\xi|f)\equiv\int d\xi_1 t_1 (\xi|\xi_1)f(\xi_1)$,
$c_1(\xi|f)\equiv\int d\xi_1 c_1(\xi|\xi_1)f(\xi_1)$.
Substituting this representation for the $*_1$--product into eq. (\ref{m13})
(or in eq. (\ref{m2Ass'})), we find
$$
f_1(\xi)[c_1(\xi|f_2f_3)-c_1(\xi|f_2)f_3(\xi)]=0,
$$
or
$$
c_1(\xi|f_1f_2)-c_1 (\xi|f_1)f_2(\xi)=0.
$$
The expression $f_1(\xi)\overleftarrow{\partial}_\alpha
\omega_1^{\alpha\beta}(\xi)\partial_\beta f_2(\xi)$ can not be
represented as $d_Ht(\xi|f_1, f_2)$. To prove this, we use the momentum
representation for the functional $t(\xi|f)$:
$$
t(\xi|f)=\sum_{k=0}t^{[\alpha]_k}(\xi)[\partial_\alpha]^kf(\xi),\quad
\varepsilon(t^{[\alpha]_k})=k\,(mod\,\,2).
$$
The summand in $d_Ht(\xi|f_1,f_2)$ with total order of the derivatives of
the functions $f_1(\xi)$ and $f_2(\xi)$ equal to 2 can arise only from the
summand in $t(\xi|f)$ with the coefficient $t^{[\alpha]_2}(\xi)$. Choosing
$$
t(\xi|f)={1\over2}t^{\alpha\beta}(\xi)\partial_\alpha\partial_\beta f(\xi),
\quad t^{\alpha\beta}=-t^{\beta\alpha},\quad\varepsilon(t^{\alpha\beta})=0,
$$
we obtain that the equality must be true
$$
f_1(\xi)\overleftarrow{\partial}_\alpha\omega_1^{\alpha\beta}(\xi)
\partial_\beta f_2(\xi)=d_Ht(\xi|f_1, f_2)=f_1(\xi)
\overleftarrow{\partial}_\alpha t^{\alpha\beta}(\xi)\partial_\beta f_2(\xi),
$$
or
$$
\omega_1^{\alpha\beta}(\xi)=t^{\alpha\beta}(\xi).
$$
The last equality is possible only for $\omega_1^{\alpha\beta}(\xi)=0$ because
$\omega_1^{\alpha\beta}(\xi)$ is a symmetric matrix while $t^{\alpha\beta}(\xi)$
is an antisymmetric one.

We have thus found that the general solution of eq. (\ref{m13})
has the form
$$
f_1*_1f_2(\xi)=m_1(\xi|f_1,f_2)=f_1(\xi)\overleftarrow{\partial}_\alpha
\omega_1^{\alpha\beta}(\xi)\partial_\beta f_2(\xi)+f_1(\xi)t_1(\xi|f_2)
-t_1(\xi|f_1f_2)+t_1(\xi|f_1)f_2(\xi)=
$$
$$
=f_1(\xi)\overleftarrow{\partial}_\alpha\omega_1^{\alpha\beta}(\xi)
\partial_\beta f_2(\xi)+d_Ht_1(\xi|f_1,f_2).
$$

If we perform the similarity transformation $T$ of the $*$--product,
$$
f_1*f_2\quad\longrightarrow\quad f_1*_{T_1}f_2=T_1^{-1}(T_1f_1*T_1f_2),
$$
with the operator\footnote{We note, that the similarity transformation $T$ of
the "pointwise" product $*_0$ with $T=1+\hbar t_1+0(\hbar^2)$ is
of the form
$$
*_{0T}=*_0+d_Ht_1+O(\hbar^2).
$$ }
$$
T_1=e^{-\hbar\hat{t}_1},
$$
then it is easy to verify that $P_{T_1}$ has the following power series
expansion in $\hbar$:
$$
P_{T_1}(\xi|f_1,f_2)=f_1(\xi)f_2(\xi)+\hbar m_1(\xi|f_1,f_2)-\hbar d_Ht_1
(\xi|f_1,f_2)+\hbar^2m _{T_12}(\xi|f_1,f_2)+O(\hbar^3)=
$$
\begin{equation}\label{m2h2}
=f_1(\xi)f_2(\xi)+\hbar f_1(\xi)\overleftarrow{\partial}_\alpha
\omega_1^{\alpha\beta}(\xi)\partial_\beta f_2(\xi)+
\hbar^2m_{T_12}(\xi|f_1,f_2)+O(\hbar^3).
\end {equation}
In what follows, we omit the index "$T_1$" at the $*_{T_1}$--product and at
the kernels $P_{T_1}$ and $m_{T_12}$.

It is now useful to introduce the $*$--commutator $[f_1, f_2]_*(\xi)$ for
two functions $f_1(\xi)$ and $f_2(\xi)$ defined by
$$
[f_1,f_2]_*(\xi)\equiv{1\over2\hbar}\biggl(f_1*f_2(\xi)-
(-1)^{\varepsilon(f_1)\varepsilon(f_2)}f_2*f_1(\xi)\biggr)
$$
and to define the Jacobian $J(\xi|f_1,f_2, f_3)$ of three functions
$f_1(\xi)$, $f_2(\xi)$ and $f_3(\xi)$ as
\begin{equation}\label{m2Jac}
J(\xi|f_1,f_2,f_3)\equiv(-1)^{\varepsilon(f_1)\varepsilon(f_3)}
[[f_1,f_2]_*f_3]_*(\xi)+\hbox{cycle}\,(1,2,3).
\end{equation}
We have
\begin{equation}\label{m2JI}
J(\xi|f_1,f_2,f_3)=0
\end{equation}
as the consequence of associativity equation (\ref{AE}) (the $*$--commutator
satisfies the Jacobi identity).

The direct calculation of the Jacobian $J(\xi|f_1,f_2,f_3)$ for $*$--product
(\ref{m2h2}) gives
$$
J(\xi|f_1,f_2,f_3)=-(-1)^{\varepsilon(f_2)+\varepsilon(f_1)\varepsilon(f_3)}
\Omega_1^{\alpha\beta\gamma}(\xi)
\partial_\alpha f_1(\xi)\partial_\beta f_2(\xi)\partial_\gamma f_3(\xi)+
O(\hbar),
$$
$$
\Omega_1^{\alpha\beta\gamma}(\xi) =
\omega^{\alpha\delta}_1(\xi)\partial_\delta\omega^{\beta\gamma}_1(\xi)
+\omega^{\gamma\delta}_1(\xi)\partial_\delta\omega^{\alpha\beta}_1(\xi)
+\omega^{\beta\delta}_1(\xi)\partial_\delta\omega^{\gamma\alpha}_1(\xi).
$$
It follows from eq. (\ref{m2JI}) that the function
$\omega^{\alpha\beta} _1(\xi)$ satisfies the Jacobi identity
$$
\omega^{\alpha\delta}_1(\xi)\partial_\delta\omega^{\beta\gamma}_1(\xi)
+\omega^{\gamma\delta}_1(\xi)\partial_\delta\omega^{\alpha\beta}_1(\xi)
+\omega^{\beta\delta}_1(\xi)\partial_\delta\omega^{\gamma\alpha}_1(\xi)=0.
$$
We note that the function $\omega^{\alpha\beta}(\xi)$ transforms as a tensor
under a change of the generators $\xi^\alpha$ $\rightarrow$
$\eta^\alpha(\xi)$. This means the following. Let operator $T_\eta$ (see eq.
(\ref{m2Change}))
describes a change of the generators. The $\hbar$ power series expansion of the
$*$--product $T_\eta^{-1} P(\xi|T_\eta f_1,T_\eta f_2)$, where
$P(\xi|f_1, f_2)$ is given by formula (\ref{m2h2}), is
\begin{equation}\label{m2Tens}
\begin{array}{c}
T_\eta^{-1}P(\xi|T_\eta f_1,T_\eta f_2)=f_1 (\xi)f_2(\xi)+
\hbar f_1(\xi)\overleftarrow{\partial}_\alpha
\omega_1^{\prime\alpha\beta}(\xi)\partial_\beta f_2(\xi)+O(\hbar^2), \\
\displaystyle\omega_1^{\prime\alpha\beta}(\xi)=\xi^\alpha
{\overleftarrow{\partial}\over\partial\zeta^\gamma}
\omega^{\gamma\delta}_1(\zeta){\partial\over\partial\zeta^\delta}\xi^\beta.
\end{array}
\end{equation}

Therefore, the tensor function $\omega^{\alpha\beta}_1(\xi)$ is the symplectic
metric of the Poisson bracket. We assume that the metric
$\omega^{\alpha\beta}_1$ is nonsingular. In this case, there exists a
change of the generators $\xi^\alpha$ $\rightarrow$ $\eta^\alpha(\xi)$
that reduces the symplectic metric to the canonical form
\begin{equation}\label{m2Can}
\omega_1^{\prime\alpha\beta}(\xi)=\lambda_\alpha\delta_{\alpha\beta},
\end{equation}
where $\lambda_\alpha=1$ in the case of the Grassman algebra over
complex numbers (${\cal G}_{{\bf C}}$) and $\lambda_\alpha=\pm1$ in the
case of the Grassman algebra over real numbers (${\cal G}_{{\bf R}}$). We
assume that the appropriate similarity transformation is performed
such that the tensor function in formula (\ref{m2h2}) is of canonical
form (\ref{m2Can}).

We represent the $*$--product as
\begin{equation}\label{m2h2'}
P(\xi|f_1,f_2)=P_G^{\hbar\lambda}(\xi|f_1,f_2)+
\hbar^2m^\prime_2(\xi|f_1,f_2)+O(\hbar^3),
\end {equation}
where
\begin{equation}\label{Gro}
P_G^{\hbar\lambda}(\xi|f_1, f_2)\equiv f_1*_{G(\hbar\lambda)}f_2(\xi)=
f_1(\xi)e^{\hbar Q_\lambda}f_2(\xi)=f_1(\xi) f_2 (\xi)+
\hbar f_1(\xi)\overleftarrow{\partial}_\alpha
\lambda_\alpha\partial_\alpha f_2 (\xi)+O(\hbar^2),
$$
$$
Q_\lambda=\sum_\alpha\lambda_\alpha\overleftarrow{\partial}_\alpha
\partial_\alpha.
\end {equation}
$P_G^{\hbar\lambda}(\xi|f_1, f_2)$ itself satisfies the associativity equation
(see Appendix). Substituting expansion (\ref{m2h2'}) into eq. (\ref{m2Ass}),
we find that the kernel $m^\prime_2$ satisfies just eq. (\ref{m2Ass'}), whence
it follows
$$
m^\prime_2(\xi|f_1,f_2)=
f_1(\xi)\overleftarrow{\partial}_\alpha\omega_2^{\alpha\beta}(\xi)
\partial_\beta f_2(\xi)+d_Ht_2(\xi|f_1, f_2)
$$
with a certain function
$\omega_2^{\alpha\beta}(\xi)=\omega_2^{\beta\alpha}(\xi)$,
$\varepsilon(\omega_2^{\alpha\beta})=0$, and a kernel $t_2(\xi|\xi_1)$,
$\varepsilon(t_2)=n\,(mod\,\,2)$. The summand $d_Ht_2(\xi|f_1,f_2)$
can be transformed out by the similarity transformation with the operator
$T=\exp{(-\hbar^2\hat{t}_2)}$. Assuming that this similarity transformation
has been already performed, we obtain that the $*$--product can be
represented as
\begin{equation}\label{m2h3}
P(\xi|f_1,f_2)=P_G^{\hbar\lambda}(\xi|f_1,f_2)+
\hbar^2f_1(\xi)\overleftarrow{\partial}_\alpha\omega_2^{\alpha\beta}(\xi)
\partial_\beta f_2 (\xi)+\hbar^3m^\prime_3(\xi|f_1,f_2)+O(\hbar^4).
\end {equation}
We assume that the $*$--product can be represented as
\begin{equation}\label{m2hk}
P(\xi|f_1,f_2)=P_G^{\hbar\lambda}(\xi|f_1,f_2)+
\hbar^kf_1(\xi)\overleftarrow{\partial}_\alpha\omega_k^{\alpha\beta}(\xi)
\partial_\beta f_2 (\xi)+\hbar^{k+1}m_{k+1}(\xi|f_1,f_2)+O(\hbar^{k+2}),
\end {equation}
$$
\omega_k^{\alpha\beta}=\omega_k^{\beta\alpha},\quad
\varepsilon(\omega_k^{\alpha\beta})=0,\quad k\ge2,
$$
after the appropriate similarity transformation.

Then the direct calculation of the Jacobian $J(\xi|f_1,f_2,f_3)$ for
this $*$--product (\ref{m2hk}) gives
$$
J(\xi|f_1,f_2,f_3)=-(-1)^{\varepsilon(f_2)+\varepsilon(f_1)\varepsilon(f_3)}
\hbar^{k-1}\Omega_k^{\alpha\beta\gamma}(\xi)
\partial_\alpha f_1(\xi)\partial_\beta f_2(\xi)\partial_\gamma f_3(\xi)+
O (\hbar^k),
$$
$$
\Omega_k^{\alpha\beta\gamma}(\xi)=
\lambda_\alpha\partial_\alpha\omega_k^{\beta\gamma}(\xi)+
\lambda_\gamma\partial_\gamma\omega_k^{\alpha\beta}(\xi)+
\lambda_\beta\partial_\beta\omega_k^{\gamma\alpha}(\xi).
$$
It follows from eq. (\ref{m2JI}) that the function
$\omega_k^{\alpha\beta}(\xi)$ satisfies the relation (the Bianchi identity)
$$
\lambda_\alpha\partial_\alpha\omega_k^{\beta\gamma}(\xi)+
\lambda_\gamma\partial_\gamma\omega_k^{\alpha\beta}(\xi)+
\lambda_\beta\partial_\beta\omega_k^{\gamma\alpha}(\xi)=0,
$$
whence we find
$$
\omega_k^{\alpha\beta}(\xi)=\lambda_\alpha\partial_\alpha\omega_k^\beta(\xi)+
\lambda_\beta\partial_\beta\omega_k^\alpha(\xi),\quad
\varepsilon(\omega_k^\alpha)=1.
$$

We perform the similarity transformation of $*$--product (\ref{m2hk})
with the  operator
$T=\exp{(-\hbar^{k-1}\omega_k^\alpha(\xi)\partial_\alpha)}$. Taking the
relations
$$
e^{-\hat{t}}\left(e^{\hat{t}}f_1(\xi)e^{\hat{t}}f_2(\xi)\right)=
f_1 (\xi) f_2(\xi),\quad\hat{t}=\omega^\alpha(\xi)\partial_\alpha,\quad
\varepsilon(\omega^\alpha)=1,
$$
$$
e^{-\hbar^k\hat{t}}P_G^{\hbar\lambda}(\xi|e^{\hbar^k\hat{t}}f_1,
e^{\hbar^k\hat{t}}f_2)=P_G^{\hbar\lambda}(\xi|f_1,f_2)+
$$
$$
+\hbar^{k+1}f_1(\xi)\overleftarrow{\partial}_\alpha
\left(\lambda_\alpha\partial_\alpha\omega^\beta(\xi)+
\lambda_\beta\partial_\beta\omega^\alpha(\xi)\right)
\partial_\beta f_2(\xi)+O(\hbar^{k+2}),\quad k\ge1,
$$
into account, we obtain that the transformed $*$--product has the form (index
$T$ is omitted)
$$
P(\xi|f_1,f_2)=P_G^{\hbar\lambda}(\xi|f_1,f_2)+
\hbar^{k+1}m^\prime_{k+1}(\xi|f_1,f_2)+O(\hbar^{k+2}),
$$
and $m'_{k+1}$ satisfies eq. (\ref{m2Ass'}). It remains to apply the
induction method.

We have thus shown that any associative $*$--product on the Grassman algebra
with boundary condition (\ref{m1BC}) can be obtained from the
$*_{G(\hbar\lambda)}$--product by a similarity transformation.

\section{Uniqueness}

We have established that any $*$--product with boundary condition
(\ref{m1BC}) and a nonsingular metric in the first--order deformation can be
reduced to the form of the $*_{G(\hbar\lambda)}$--product by the similarity
transformations generated by the operators of the form
\begin{equation}\label{m2Trans}
T=T_\eta(1+\hbar t_1+O(\hbar^2)),
\end{equation}
$T_\eta$ is the operator of a nonsingular change of the generators. It is
essential that it is impossible to reduce the $*$--product with a nonsingular
tensor $\omega_1^{\alpha\beta}(\xi)$ to the "pointwise" product by a
similarity transformations of this type because $t_1$ does not contribute
to $\omega_1^{\alpha\beta}(\xi)$ and $T_\eta$ does not violate the
nonsingularity of the tensor $\omega_1^{\alpha\beta}(\xi)$ (see eq.
(\ref{m2Tens})). The question arises whether there exists a similarity
transformation generated by an operators $T$ of the more general form that
does reduce the $*_{G(\hbar\lambda)}$--product to the "pointwise" one?
We now show that the answer to this question is in negative.

In fact, any nonsingular operator $T$ can be written as
$$
T=T'(1+O(\hbar)),
$$
with a certain nonsingular operator $T'=T\big|_{\hbar=0}$. By virtue of the
boundary condition for the $*$--product (see eqs. (\ref{m1Ser}) and
(\ref{m1BC})) the $T'$--similarity transformation should leave the
"pointwise" product invariant, i.e., the operator $T'$
should satisfy the equation
\begin{equation}\label{m2Inv}
T'(\xi|f_1)T'(\xi|f_2)=T'(\xi|f_1f_2).
\end{equation}
We now show that the general solution of eq. (\ref{m2Inv}) is
a nonsingular change of the generators
$$
T'(\xi|f)=T_\eta(\xi|f)=f(\eta(\xi)),
$$
with some functions $\eta^\alpha(\xi)$, $\varepsilon(\eta^\alpha)=1$.

Let $f_2(\xi)=1$, then it follows from (\ref{m2Inv}) that
$$
T'(\xi|f_1)(1-T'(\xi|1))=0\quad\Longrightarrow\quad T'(\xi|1)=\int d\xi'
T'(\xi |\xi')=1
$$
because $T'$ is assumed to be nonsingular.

We now choose $f_1(\xi)=\exp{(\xi^\alpha p_\alpha)}$, $f_2(\xi)=
\exp{(\xi^\alpha q_\alpha)}$, where $p_\alpha$ and $q_\alpha$ are odd
generators (in the wider Grassman algebra with the generators $\xi^\alpha$,
$p^\alpha$, $q^\alpha$), and introduce the notation
$$
\int d\xi f(\xi)e^{\xi^\alpha p_\alpha}\equiv\tilde{f}(p).
$$
It follows from relation (\ref{m2Inv}) that
$$
\tilde{T}'(\xi|p)\tilde{T}'(\xi|q)=\tilde{T}'(\xi|p+q),\quad
\tilde{T}'(\xi|0)=1,
$$
whence we obtain
$$
\tilde{T}'(\xi|p)=e^{\eta_\alpha(\xi)p_\alpha},\quad
\eta^\alpha(\xi)\equiv\tilde{T}'(\xi|p)
{\overleftarrow\partial\over\partial p_\alpha}\biggl|_{p=0}.
$$
The inverse Fourier transform of $\tilde{T}'(\xi|p)$ is
$$
T'(\xi|\xi')=\delta(\eta(\xi)-\xi').
$$

Thus, we have proved the following.

{\bf Theorem}

Any associative $*$--product on the Grassman algebra ${\cal G}_K$ with
boundary condition (\ref{m1BC}) and the nonsingular symplectic metric
in the first order deformation $*_1$ is equivalent to
(i.e., is related by a similarity transformation to) the $*$--product
$P_G^{\hbar\lambda}(\xi|f_1,f_2)$, which  is not equivalent
to the "pointwise" product for $\hbar\neq0$,. We also note that the
$*_{G(\hbar\lambda)}$--products with different values of the parameter
$\hbar$ can be related by a similarity transformation (generated by a scale
transformation of the generators $\xi^\alpha$).

\section{Relation to the Clifford algebra}

In this Section, we discuss the relation of the algebra ${\cal F}_K$, the
algebra of the elements of the Grassman algebra
${\cal G}_K$ (i.e., of the functions of the generators) with the $*$--product
as a product, to the Clifford algebra ${\cal K}_K$ over the field $K$.

We take the $*_{G(\lambda)}$--product\footnote{In this Section, we choose
$\hbar=1$ that is equivalent to the transition to the new generators
$\xi^{\prime\alpha}=\sqrt\hbar\xi^\alpha$} for a product in the algebra
${\cal F}_K$. The comment on the general $*$--product is at
the end of the Section.

It is easy to verify that the $*_{G(\lambda)}$--product has the following
properties:
\begin{equation}\label{Prop1}
c*_{G(\lambda)}f(\xi)=f*_{G(\lambda)}c(\xi)=cf(\xi)
\end{equation}
for any function $f(\xi)$ and $c=const$,
\begin{equation}\label{Prop2}
\xi^{\alpha_1}*_{G(\lambda)}\xi^{\alpha_2}*_{G(\lambda)}\cdots*_{G(\lambda)}
\xi^{\alpha_k}=\xi^{\alpha_1}\xi^{\alpha_2}\cdots\xi^{\alpha_k},\quad
\alpha_1<\alpha_2<\cdots<\alpha_k.
\end{equation}
The explicit calculation gives
$$
\xi^\alpha*_{G(\lambda)}\xi^\beta=\xi^\alpha\xi^\beta+
\lambda_\alpha\delta_{\alpha\beta},
$$
therefore, we have
\begin{equation}\label{Prop3}
\xi^\alpha*_{G(\lambda)}\xi^\beta+\xi^\beta*_{G(\lambda)}\xi^\alpha=
2\lambda_\alpha\delta_{\alpha\beta},
\end{equation}
i.e., the generators $\xi^\alpha$ form the Clifford algebra with the
$*_{G(\lambda)}$--product.

Collecting together definition (\ref{Bas}) and properties
(\ref{Rep}), (\ref{Prop1}), (\ref{Prop2}), and (\ref{Prop3}), we obtain that
the following Statement is true:

The algebra ${\cal F}_K$ with the $*_{G(\lambda)}$--product as a
product is isomorphic to the Clifford algebra ${\cal K}_K$
over $K$ with the generators $\gamma^\alpha$, $\gamma^\alpha\gamma^\beta+
\gamma^\beta\gamma^\alpha=2\lambda_\alpha\delta_{\alpha\beta}$. The
isomorphism ${\cal F}_K$ $\Longleftrightarrow$ ${\cal K}_K$ is given by
\begin{equation}\label{Iso}
{\cal F}_K\ni f(\xi)=\sum_If_Ie^I\quad\Longleftrightarrow\quad
K_f=\sum_If_I\Gamma^I\in{\cal K}_K,
\end{equation}
$$
f_1*_{G(\lambda)}f_2(\xi)\quad\Longleftrightarrow\quad K _ {f_1} K _ {f_2},
$$
where
$$
\Gamma^0=1,\quad\Gamma^\alpha=\gamma^\alpha,\quad
\Gamma^{\alpha_1\alpha_2\cdots\alpha_k}=
\gamma^{\alpha_1}\gamma^{\alpha_2}\cdots\gamma^{\alpha_k},\quad k\ge2.
$$

We note that there is the natural $Z_2$--grading $g$ in this algebra, namely,
$g(f)$, where $f$ is considered an element of the algebra
${\cal F}_K$, coincides with the Grassman parity $\varepsilon(f)$, where $f$
is considered an element of the algebra ${\cal G}_K$. If we introduce the
$Z_2$--grading $g$ in the Clifford algebra,
$$
g(\Gamma^I)=k\,(mod\,\,2),\quad I=\alpha_1\cdots\alpha_k,
$$
then isomorphism (\ref{Iso}) preserves the grading.

What we can say about the representations of the algebra
${\cal F}_K$, which are simultaneously the representations of
the Clifford algebra? It is natural to consider the exact representations.

As is well known, the algebras ${\cal K}_{\bf C}$ and ${\cal K}_{\bf R}$
of even dimension $n=2m$ and the algebra ${\cal K}_{\bf R}$ of odd
dimension $n=2m+1$ with $\delta\lambda_{(n)}\equiv
\sum\limits_\alpha\lambda_\alpha=3\,(mod\,\,4)$\footnote{The property
$(\Gamma^{12\cdots n})^2=-1$ holds for this algebra ${\cal K}_{\bf R}$.} have
only one
irreducible representation (for fixed $n$ and $\delta\lambda_{(n)}$), the
representation being exact\footnote{The matrices realizing the basis
$\Gamma^I$ are linearly independent.}. In these cases, the algebra
${\cal F}_K$ has a unique exact irreducible representation coinciding with the
irreducible representation of the algebra ${\cal K}_K$.

The algebra ${\cal K}_{\bf C}$ of odd dimension $n=2m+1$ and the algebra
${\cal K}_{\bf R}$ of odd dimension $n=2m+1$ and
$\delta\lambda_{(n)}=1\,(mod\,4)$\footnote{The property
$(\Gamma^{12\cdots n})^2=1$ holds for this algebra ${\cal K}_{\bf R}$.} are
decomposed into a direct sum of two simple subalgebras. One of these
subalgebras is realized by zero in the irreducible representations. Therefore
the irreducible representations of the Clifford algebra can not be the exact
representations of the algebra ${\cal F}_K $ (and of the algebra
${\cal K}_K$). The minimum exact representation $V$ of the algebra
${\cal F}_K $ is decomposed into a
direct sum of two irreducible nonequivalent representations, $V=V_++V_-$,
the projectors $P_\pm$ on $V_{\pm}$ are $P_\pm=(1\pm\Gamma^{12\cdots n})/2$.

Finally, we consider the general $*$--product. Let $T$ be
the operator generating the similarity transformation of the $*$--product
to the $*_{G(\lambda)}$--product:
$$
f_1*f_2=T^{-1}\biggl(Tf_1*_{G(\lambda)}Tf_2\biggr).
$$
Then the isomorphism ${\cal F}_K$ $\Longleftrightarrow$ ${\cal K}_K$ is given
by
\begin{equation}\label{IsoT}
{\cal F}_K\ni f(\xi)=\sum_If_Ie^I\quad\Longleftrightarrow\quad K_f
=\sum_If_IT^I_J\Gamma^J\in{\cal K}_K,
\end {equation}
$$
f_1*f_2(\xi)\quad\Longleftrightarrow\quad K_{f_1}K_{f_2},
$$
where the matrix $T^I_J$ is defined by
$$
Te^I=T^I_Je^J.
$$
Isomorphism (\ref{IsoT}) of the algebras ${\cal G}_K$ and
${\cal K}_K$ preserves the above--introduced grading because we consider
only even operators $T$.

{\bf Acknowledgments}

The author thanks Tipunin I. and Voronov B. for useful discussions and RFBR,
contract 99--02--17916 and school--contract 00--15--96566, for support.

\def\theequation{A.\arabic{equation}}
\setcounter{equation}{0}

\section*{Appendix}

For completeness, we prove the statement that $*_{G(\lambda)}$--product
(\ref{Gro}) with arbitrary $\lambda_\alpha$ ($\hbar=1$) satisfies
associativity equation (\ref{AE}). In what follows, we omit the bottom index
``$G(\lambda)$'' and write the top index ``$(n)$'' at the $*$--product
symbols, where $n$ is the number of the generators of the Grassman
algebra. Similarly, we also omit the bottom index ``$\lambda$'' and write the
top index ``$(n)$'' at the operator $Q$ in the definition of the
$*_G$--product. The operator $Q^{(n)}$ satisfies the relations
$$
Q^{(n)}=Q^{(n-1)}+\lambda_n\overleftarrow{\partial}_n\partial_n,
$$
$$
e^{Q^{(n)}}=(1+\lambda_n\overleftarrow{\partial}_n\partial_n)e^{Q^{(n-1)}}.
$$

We prove the statement by induction in the number $n$ of the generators.

Let $\varphi(\xi)$ and $\bar{\varphi}(\xi)$ be the elements of the
Grassman algebra that are independent of the generator $\xi^n$:
$$
{\partial\over\partial\xi^n}\varphi(\xi)\equiv0,\quad
{\partial\over\partial\xi^n}\bar{\varphi}(\xi)\equiv0.
$$
Any element $f(\xi)$ can be represented as
$$
f(\xi)=\varphi(\xi)+\xi^n\bar{\varphi}(\xi),\quad
\varepsilon(\varphi)=\varepsilon(f),\quad
\varepsilon(\bar{\varphi})=\varepsilon(f)+1\,(mod\,\,2).
$$
The $*^{(n)}$--product is represented in terms of $\varphi$ and
$\bar{\varphi}$ as follows:
$$
f_1*^{(n)}f_2(\xi)=\varphi_{12}(\xi)+\xi^n\bar{\varphi}_{12}(\xi),
$$
$$
\varphi_{12}(\xi)=\varphi_1*^{(n-1)}\varphi_2(\xi)+
\lambda_n(-1)^{\varepsilon(\bar{\varphi}_1)}
\bar{\varphi}_1*^{(n-1)}\bar{\varphi}_2(\xi),
$$
$$
\bar{\varphi}_{12}(\xi)=\bar{\varphi}_1*^{(n-1)}\varphi_2(\xi)+
(-1)^{\varepsilon(\varphi_1)}\varphi_1*^{(n-1)}\bar{\varphi}_2(\xi).
$$
The associator of the $*^{(n)}$--product is
$$
(f_1*^{(n)}f_2)*^{(n)}f_3(\xi)-f_1*^{(n)}(f_2*^{(n)}f_3)(\xi)=\varphi(\xi)+
\xi^n\bar{\varphi}(\xi),
$$
$$
\begin{array}{r}
\varphi=\varphi_{(12)3}-\varphi_{1(23)}=
[(\varphi_1*^{(n-1)}\varphi_2)*^{(n-1)}\varphi_3-
\varphi_1*^{(n-1)}(\varphi_2*^{(n-1)}\varphi_3)]+ \\
+\lambda_n(-1)^{\varepsilon(\bar{\varphi}_1)}
[(\bar{\varphi}_1*^{(n-1)}\bar{\varphi}_2)*^{(n-1)}\varphi_3-
\bar{\varphi}_1*^{(n-1)}(\bar{\varphi}_2*^{(n-1)}\varphi_3)]+ \\
+\lambda_n(-1)^{\varepsilon(\bar{\varphi}_1)+\varepsilon(\varphi_2)}
[(\bar{\varphi}_1*^{(n-1)}\varphi_2)*^{(n-1)}\bar{\varphi}_3-
\bar{\varphi}_1*^{(n-1)}(\varphi_2*^{(n-1)}\bar{\varphi}_3)]+ \\
+\lambda_n(-1)^{\varepsilon(\bar{\varphi}_2)}
[(\varphi_1*^{(n-1)}\bar{\varphi}_2)*^{(n-1)}\bar{\varphi}_3-
\varphi_1*^{(n-1)}(\bar{\varphi}_2*^{(n-1)}\bar{\varphi_3})]\phantom{a},
\end{array}
$$
$$
\begin{array}{r}
\bar{\varphi}=\bar{\varphi}_{(12)3}-\bar{\varphi}_{1(23)}=
[(\bar{\varphi}_1*^{(n-1)}\varphi_2)*^{(n-1)}\varphi_3-
\bar{\varphi}_1*^{(n-1)}(\varphi_2*^{(n-1)}\varphi_3)]+ \\
+(-1)^{\varepsilon(\varphi_1)}
[(\varphi_1*^{(n-1)}\bar{\varphi}_2)*^{(n-1)}\varphi_3-
\varphi_1*^{(n-1)}(\bar{\varphi}_2*^{(n-1)}\varphi_3)]+ \\
+(-1)^{\varepsilon(\varphi_1)+\varepsilon(\varphi_2)}
[(\varphi_1*^{(n-1)}\varphi_2)*^{(n-1)}\bar{\varphi}_3-
\varphi_1*^{(n-1)}(\varphi_2*^{(n-1)}\bar{\varphi}_3)]+ \\
+\lambda_n(-1)^{\varepsilon(\bar{\varphi}_2)}
[(\bar{\varphi}_1*^{(n-1)}\bar{\varphi}_2)*^{(n-1)}\bar{\varphi}_3-
\bar{\varphi}_1*^{(n-1)}(\bar{\varphi}_2*^{(n-1)}
\bar{\varphi_3})]\phantom {a}.
\end{array}
$$
We can see from these formulas that if the $*^{(n-1)}$--product is
associative, then the $*^{(n)}$--product is also associative. In addition, it
is easy to see from the same formulas that the $*^{(1)}$--product is
associative (all $\varphi_i$ and $\bar{\varphi}_i$ are numbers,
$\varepsilon(\varphi_i)=\varepsilon(\bar{\varphi}_i)=0$ and the
$*^{(0)}$--product is the usual product of numbers). $\Box$

\end{document}